\documentclass[aps,prx,superscriptaddress,twocolumn]{revtex4-2}

\usepackage[normalem]{ulem}
\usepackage{graphicx}
\usepackage{xcolor}
\usepackage{amsmath}
\usepackage{amssymb}
\usepackage{siunitx}
\DeclareSIUnit{\sqrthz}{\ensuremath{\sqrt{\mathrm{Hz}}}}

\usepackage[english]{babel}
\usepackage{booktabs}

\usepackage[colorlinks=true, pdfstartview=FitV, linkcolor=blue,
citecolor=blue, urlcolor=blue]{hyperref}
\usepackage[capitalise,nameinlink]{cleveref}

\begin{document}
\title{Scanning nanowire quantum dot with radio-frequency read-out}
\author{A. Kamber}
\thanks{These authors contributed equally to this work.}
\affiliation{Department of Physics, University of Basel, 4056 Basel, Switzerland}

\author{L. Forrer}
\thanks{These authors contributed equally to this work.}
\affiliation{Department of Physics, University of Basel, 4056 Basel, Switzerland}
\affiliation{Swiss Nanoscience Institute, University of Basel, 4056 Basel, Switzerland}

\author{V. Zannier}
\affiliation{NEST, Istituto Nanoscienze-CNR and Scuola Normale Superiore, I-56127 Pisa, Italy}

\author{F. Beltram}
\affiliation{NEST, Istituto Nanoscienze-CNR and Scuola Normale Superiore, I-56127 Pisa, Italy}

\author{L. Sorba}
\affiliation{NEST, Istituto Nanoscienze-CNR and Scuola Normale Superiore, I-56127 Pisa, Italy}

\author{M. Poggio}
\affiliation{Department of Physics, University of Basel, 4056 Basel, Switzerland}
\affiliation{Swiss Nanoscience Institute, University of Basel, 4056 Basel, Switzerland}

\author{F. R. Braakman}
\email[Corresponding author: ]{floris.braakman@unibas.ch}
\affiliation{Department of Physics, University of Basel, 4056 Basel, Switzerland}
\affiliation{Swiss Nanoscience Institute, University of Basel, 4056 Basel, Switzerland}

\begin{abstract}
\noindent Scanning single-electron transistors (SETs) are highly sensitive local probes of charge and electric potential, but their integration with robust height control for scanning probe microscopy remains challenging. Furthermore, readout bandwidths are limited to tens of kHz, while characterizing faster charge dynamics is often of interest. Here, we present a scanning SET in the form of a nanowire (NW) quantum dot (QD) integrated at the apex of a non-contact atomic force microscopy (NC-AFM) cantilever. We demonstrate NC-AFM topographic imaging and local electrometry capability by mapping the electrostatic potential generated by a set of nanoscale voltage-biased gate electrodes. Comparison with electrostatic simulations shows agreement with the measured potential profiles. We demonstrate  radio-frequency (rf) reflectometry read-out of the sensor, by embedding the QD in an LC-resonator. This platform establishes a route towards high-bandwidth scanning electrometry with simultaneous topographic contrast.
\end{abstract}
\maketitle

\section*{Introduction}
Sensitive electrometers play an important role in exploring the electronic properties of the growing variety of quantum materials and nanoscale devices. Scanning single-electron transistors are among the most sensitive electrometers available \cite{devoretAmplifyingQuantumSignals2000}, with voltage and charge sensitivities down to a few $\mathrm{\mu}\textrm{V}\sqrt{\textrm{Hz}}^{-1}$ and $\mathrm{\mu}e\sqrt{\textrm{Hz}}^{-1}$, respectively. Since the first scanning SET introduced by Yoo et al. \cite{yooScanningSingleElectronTransistor1997a}, the combination of high charge sensitivity with the spatial control of scanning probe microscopy has enabled a variety of local studies of e.g. the quantum Hall effect \cite{yacobyElectricalImagingQuantum1999,zhitenevImagingLocalizedElectronic2000,ilaniMicroscopicNatureLocalization2004,martinNatureLocalizationGraphene2009, feldmanUnconventionalSequenceFractional2012}, different regimes of electron flow\cite{ellaSimultaneousVoltageCurrent2019,sulpizioVisualizingPoiseuilleFlow2019,kumarImagingHydrodynamicElectrons2022}, and various correlated electronic systems \cite{zondinerCascadePhaseTransitions2020a, rozenEntropicEvidencePomeranchuk2021, pierceUnconventionalSequenceCorrelated2021, yuCorrelatedHofstadterSpectrum2022, yuSpinSkyrmionGaps2023}.
These experiments illustrate the power of imaging local electronic properties with high sensitivity and low invasiveness, a capability equally relevant to the development of devices for quantum computation\cite{marchioriMagneticImagingSuperconducting2022, denisovMicrowaveFrequencyScanningGate2022, denisovDispersiveReadoutSilicon2023a}.
The reach of scanning electrometry, however, remains limited by the probes themselves. SETs fabricated on tapered glass tips or on rigid beams lack a sensitive topographic channel, making tip–sample distance control during scanning difficult and restricting their use to samples with essentially flat surfaces. In addition, readout is typically confined to dc or low-frequency transport, whose limited bandwidth precludes imaging of charge dynamics and leaves measurements exposed to ubiquitous low-frequency noise.

Scanning probe microscopy based on non-contact atomic force-microscopy cantilevers allows for controlled tip–sample approach, navigation, and topographic imaging. This enables a safe approach without risking damage to the fragile SET, permits scanning across complex topographies with large height variations - such as quantum computing devices with dense gate electrodes and other on-chip structures - and speeds up locating the region of interest. Although several approaches towards developing SETs on force microscopy cantilevers have been reported \cite{brenningSingleElectronTransistor2006, suSensitiveChargeScanning2016}, none of these have been demonstrated as scanning electrometry probes.

\begin{figure*}[t]
	\centering
	\includegraphics[width=\textwidth]{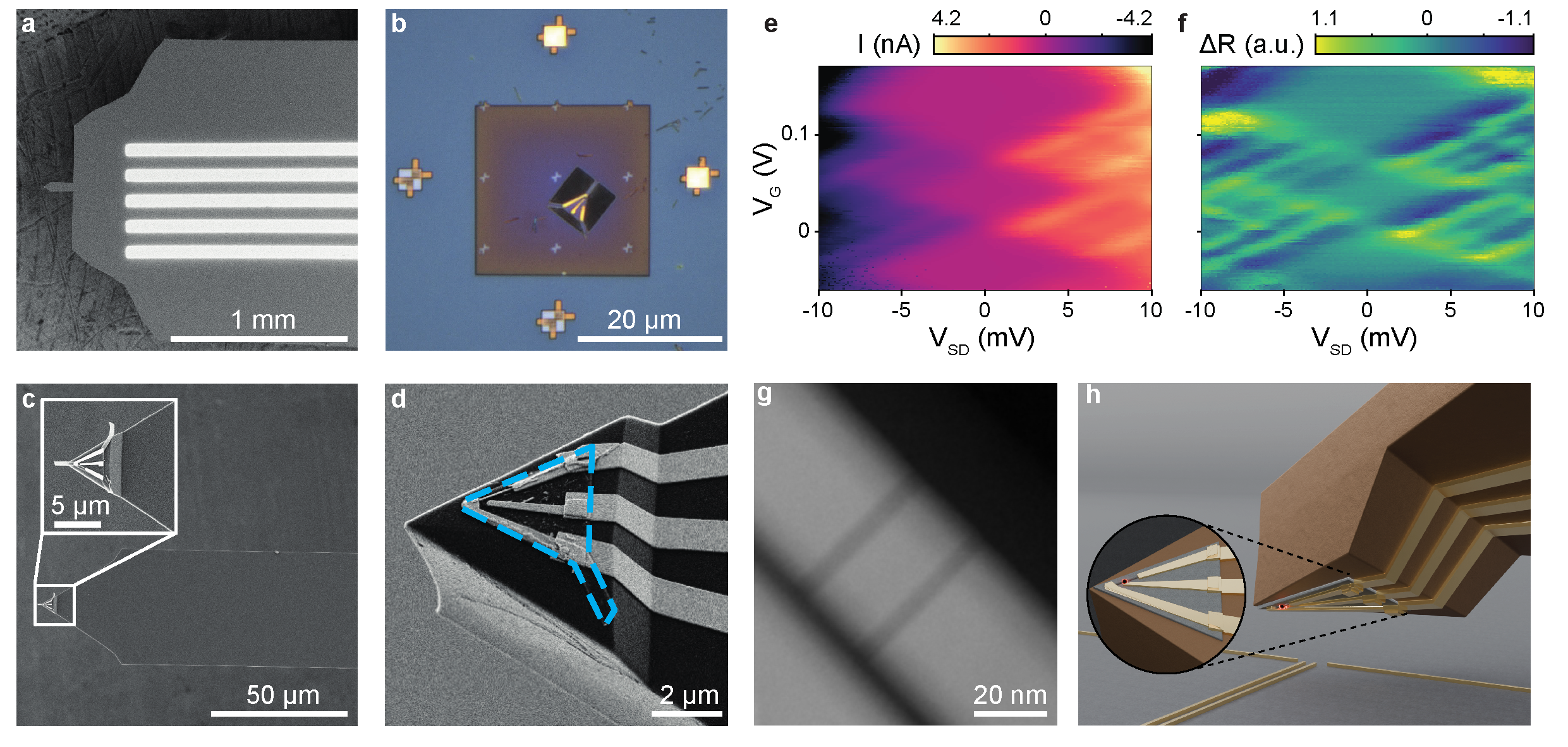}
	\caption{\textbf{Overview of the probe fabrication and characterization.} \textbf{a} Scanning electron micrograph of a highly doped Si cantilever, coated with $\sim$\SI{60}{\nano\meter} of AlO$_x$. On the cantilever chip, five Ti/Au bond pads defined through shadow-mask evaporation are visible. \textbf{b} Optical microscope image of a \SI{50}{\nano\meter} thick SiN membrane (square), suspended from a Si frame. The membrane hosts a contacted and gated InAs/InP NW QD, separated from the surrounding membrane leaving only three narrow scaffolding beams in a dry etching step. \textbf{c} Scanning electron micrograph of a NC-AFM cantilever with the triangular SiN chip containing the prefabricated device shown in \textbf{b} transferred to the flat mesa prepared at the cantilever tip and subsequently cleaned in an oxygen plasma. \textbf{d} Scanning electron micrograph of the cantilever apex, with the sensor on its SiN chip (outlined by dashed blue lines) electrically connected to the bond pads on the cantilever chip. \textbf{e, f} Coulomb diamond measurements of sensor QD. \textbf{e} dc measurement. \textbf{f} Corresponding transconductance measurement obtained by lock-in detection at 1.25 kHz. \textbf{g} Scanning transmission electron micrograph of an InAs NW (light) with axially grown InP tunnel barriers (dark). \textbf{h} Sketch of scanning probe setup with SNWQD over sample with gate electrodes.}
	\label{fig:figure_1}
\end{figure*}
Here, we present a versatile scanning probe platform in which a gated and electrically contacted quantum dot is integrated at the apex of a force microscopy cantilever, combining sensitive charge detection with NC-AFM topographic contrast. We describe a fabrication procedure that allows for on-cantilever patterning of multiple gates and contacts of arbitrary shape and that is not tied to a particular QD system: nanowires, carbon nanotubes, or planar material can in principle be integrated in the same way, with contacting and gating schemes tailored to the application. 
Furthermore, the platform supports rf reflectometry readout\cite{reillyFastSinglechargeSensing2007}, which offers substantially increased bandwidth over dc and low-frequency techniques and is therefore attractive for probing fast charge dynamics and other non-equilibrium phenomena\cite{vigneauProbingQuantumDevices2023}. Because each cantilever electrode is individually addressable, the reflectometry circuitry can be coupled either to an ohmic contact or to a gate, the latter enabling dispersive readout of the sample in combination with a complex potential landscape imprinted by the remaining probe gates.
Among the possible QD systems, NW QDs are particularly well suited for scanning probes. Their small size ($\sim$10–50 nm) and the natural one-dimensional confinement provided by the NW geometry mean that few gates are required for QD control, and the slender NW allows the QD to be brought to within close vicinity of a sample surface. Importantly, the strong confinement and tunability of NW QDs endow them with a higher sensitivity to electric fields and nearby charges than conventional metallic SETs, whilst remaining functional at temperatures above 4 K. 

In the demonstration presented here, the QD is defined between two axial InP tunnel barriers embedded in an InAs NW\cite{zannierGrowthDynamicsInAs2019,thomasHighlySymmetricTunable2020}. The as-grown tunnel barriers allow the QD to be operated with a single gate and using relatively small gate voltages, so that the probe can be tuned to its most sensitive point while remaining minimally invasive to the sample under study. The NW QD device is placed at the apex of a prepared cantilever tip, electrically contacted, and gated. We demonstrate the functionality of the probe by recording topographic images of a four-gate structure and by quantitatively mapping the three-dimensional electrostatic potential generated by voltages applied to the gates. Finally, we implement rf reflectometry by embedding the scanning SET in an LC-resonator and characterize the device in the rf domain, establishing that the same probe is compatible with both scanning electrometry and high-bandwidth rf readout. Although the sensitivity and bandwidth achieved in this initial implementation are still modest, they are likely limited by the unoptimized high-frequency environment of the cantilever rather than by the sensor itself.

\section*{Scanning Nanowire Quantum Dot Fabrication}
A key feature of our approach is that the sensor is fabricated onto a finished
cantilever, making the process compatible with fragile device architectures,
such as NW QDs, that may not survive cantilever fabrication. \Cref{fig:figure_1} illustrates the key steps of the fabrication process, which
is described in detail in Appendix I. A commercially available
NC-AFM cantilever (Nanosensors ATEC-NC) is prepared by focused-ion-beam milling a flat mesa at its apex, followed by the deposition of an insulating
AlO$_x$ layer and of Ti/Au bond pads on the cantilever chip
(\cref{fig:figure_1}\textbf{a}). The QD device, employing an InAs/InP nanowire (see \cref{fig:figure_1}\textbf{g}), is fabricated
separately on a suspended silicon nitride membrane using conventional electron
beam lithography, with source and drain contacts and a gate adjacent
to the QD. A subsequent etching step releases the device from the membrane
except for narrow sacrificial scaffolding beams (\cref{fig:figure_1}\textbf{b}). This
membrane-based route allows a complete, prefabricated device (rather than an
individual nanowire) to be transferred to the cantilever, and provides
mechanical stability during processing.
\begin{figure*}[t]
	\centering
	\includegraphics[width=\textwidth]{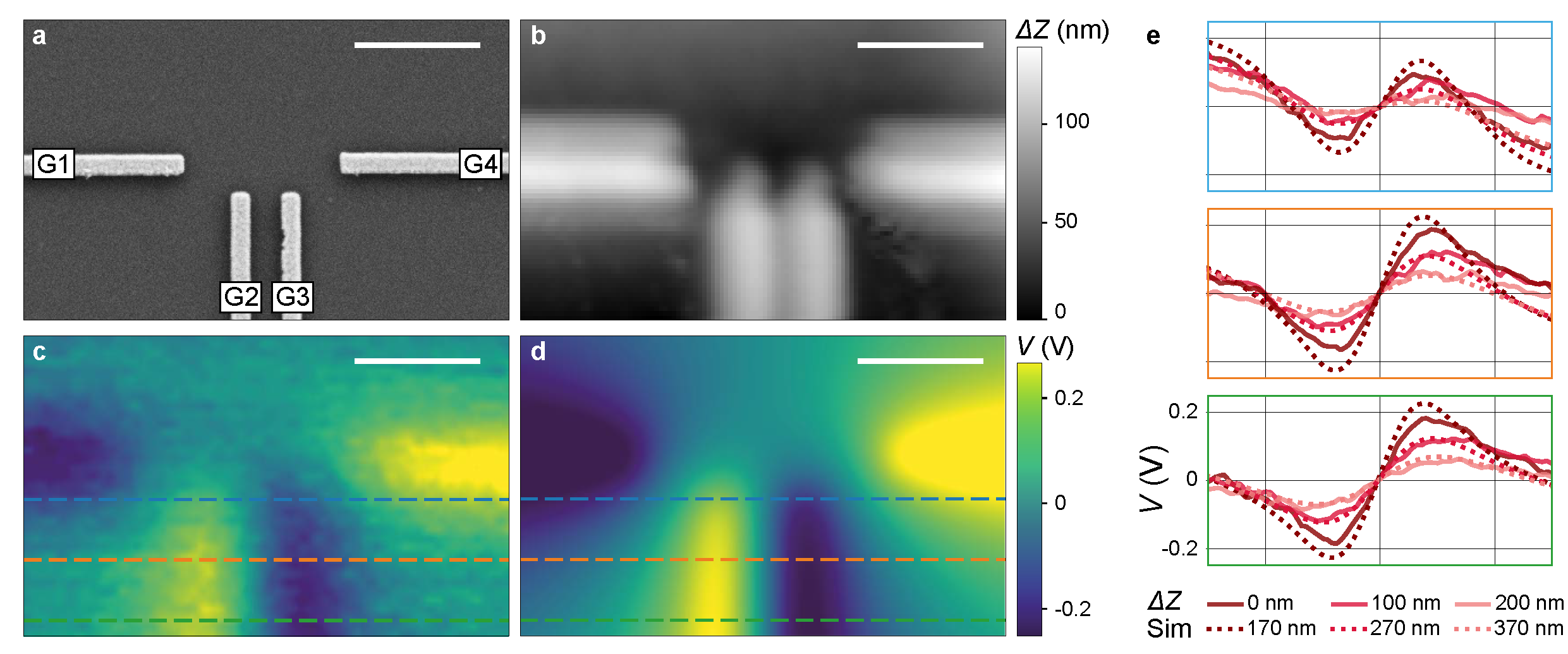}
	\caption{\textbf{Imaging the electrostatic potential of nanoscale gate electrodes.} \textbf{a} Scanning electron micrograph of the gate structure investigated in the experiment. \textbf{b} NC-AFM topography image of the gates shown in \textbf{a}, recorded at constant frequency shift. The scale represents the relative height $\Delta Z$, normalized to a minimum of zero since the absolute scanning height is undetermined. \textbf{c} Image of electrostatic potential generated by biasing gates G1 and G3 (G2 and G4) with \SI{-0.5}{\volt} (\SI{0.5}{\volt}). During scanning, a slow feedback loop maintains the QD at a constant operating point by compensating the electrostatic potential generated by the underlying gates through the voltage applied to the QD gate. The recorded compensating voltage is converted to the electrostatic potential at the QD using the lever arm $\alpha = C_{\mathrm{G}}/C_{\mathrm{\Sigma}}$. The three colored dashed lines indicate the horizontal linecuts used for comparison in \textbf{e}. \textbf{d} Electrostatic simulation of the investigated gate structure with applied voltages of $\pm$\SI{0.5}{\volt}, calculated using COMSOL. \textbf{e} Comparison between simulation and experiment for the three linecuts indicated in \textbf{c} and \textbf{d}. Dashed curves correspond to the simulation and solid curves to the measured data. All scale bars represent \SI{500}{\nano\meter}.}
	\label{fig:figure_2_gate_scans}
\end{figure*}
Using a micromanipulator setup, the prefabricated device is transferred to the
cantilever mesa, breaking the scaffolding beams in the process
(\cref{fig:figure_1}\textbf{c}). The device is positioned with the QD close to the
front edge of the mesa to minimize QD-sample distance and to maximize electrostatic coupling to the scanned
structures and thereby the sensitivity. In a final electron beam lithography
step, based on a floating-resist coating technique
\cite{forrerElectronbeamLithographyNanostructures2023}, the device contacts are connected to the bond pads
on the cantilever chip, yielding the completed probe shown in
\cref{fig:figure_1}\textbf{d}. Although demonstrated here for InAs/InP nanowires, the
approach is directly compatible with a broad range of nanowire-based devices and could
be extended to routinely fabricated planar sensors.\\

\section*{Scanning Nanowire Quantum Dot Characterization}
We evaluate the sensitivity of the sensor QD using typical electrical transport methods (see further Appendix III). These measurements are performed using the custom-built scanning probe setup mounted inside a liquid-helium bath cryostat that is later also used for demonstrating imaging capabilities. 
Figs.\ref{fig:figure_1}\textbf{e} and \textbf{f} present measurements of the QD source-drain current and the transconductance (\SI{1.25}{\kilo\hertz}, \SI{2.2}{\milli\volt} rms excitation amplitude on the QD gate), respectively, as a function of source-drain bias $V_\mathrm{SD}$ and gate voltage $V_{\mathrm{G}}$. The measurements exhibit a characteristic Coulomb diamond pattern\cite{thomasHighlySymmetricTunable2020}, confirming that the QD remains functional after the transfer, etching, and lithography processes. Note further that Coulomb oscillations already appear at gate voltages close to zero, enabling low gate voltage invasiveness. We use such Coulomb diamond measurements to determine the electron temperature to be 6.2$\pm$1.3 K and the lever arm to be 8.9$\pm$1.8 \%, where the uncertainties arise from the statistical fitting error and from the estimation of the Coulomb diamond slopes and the peak-to-peak spacing $\Delta V_{\mathrm{G}}$.
Next we use a measurement of the voltage noise density at \SI{1.25}{\kilo\hertz}, recorded with the probe retracted from the sample, to determine the voltage sensitivity $\delta V$ and induced charge sensitivity $\delta Q = \delta V C_{\mathrm{G}}$. Here $C_{\mathrm{G}}$ is the capacitance of the QD gate to the QD. Dividing the voltage noise density by the system gain obtained from the dc slope $\partial I/\partial V_{\mathrm{G}}$ of a Coulomb peak, we find $\delta V =$ \SI{4.5e-6}{\volt\per\sqrthz} and $\delta Q=$ \SI{65e-6}{\elementarycharge\per\sqrthz}. Furthermore, dividing the same voltage noise density by the sensor response extracted from a lock-in measurement at \SI{1.25}{\kilo\hertz}, we find a voltage sensitivity of \SI{30e-6}{\volt\per\sqrthz} and a corresponding charge sensitivity of \SI{425e-6}{\elementarycharge\per\sqrthz}. Compared to the dc-slope-derived limit, this suppressed experimental response can likely be attributed to the RC low-pass filter formed by the QD resistance and parasitic capacitances. This limitation can in principle be overcome by employing QD readout based on rf-reflectometry, as discussed later. 

\section*{Electrostatic potential imaging}
We next demonstrate the ability of our scanning NW QD to perform quantitative imaging of the electric potential using a sample with four nanoscale gate electrodes. The Ti/Pd/Au (thicknesses 5/20/50 nm) gate electrodes have a design width of \SI{100}{\nano\meter} and are defined on top of a silicon substrate with a \SI{300}{\nano\meter} thick thermal oxide. An SEM image of the gate pattern is shown in \cref{fig:figure_2_gate_scans}\textbf{a}. We perform the imaging using the home-built scanning probe setup mentioned above, which is housed under high vacuum inside a 4He bath cryostat. The setup includes a stack of piezoelectric positioners to move the sample with respect to the fixed cantilever, allowing for close tip-sample approach and for lateral scanning. The cantilever's displacement is measured via a fiber-optic interferometer at a wavelength of \SI{1550}{\nano\meter} and incident power of \SI{1}{\micro\watt}. The interferometer is focused on the cantilever top surface using a second stack of piezoelectric positioners. For topographic contrast, we track the frequency of the fundamental flexural mode of the cantilever, which has a resonance frequency of \SI{331}{\kilo\hertz} and a quality factor of 15,000 in high vacuum and at a measured temperature of \SI{4.8}{\kelvin} of the scanning probe microscope. For more information on the scanning probe setup, as well as the cantilever printed circuit board (PCB) used to electrically connect to the SET, see Appendices IV and II, respectively.

To image the electrostatic potential generated by the biased sample gates, we first record an NC-AFM topography image at a constant frequency shift as shown in \cref{fig:figure_2_gate_scans}\textbf{b}. Topographic resolution is limited by the size of the mesa at the tip apex, as well as the SET structure patterned on it. The height map is used to determine the highest point of the scan area and allows for scans at constant height. As we observed that using the laser from the interferometric cantilever displacement readout results in increased current noise through the QD, the laser is switched off during the following constant height scans.

During scanning, we apply a source-drain bias of \SI{1}{\milli\volt} across the QD and an ac excitation of \SI{40}{\milli\volt} at \SI{131.15}{\hertz} to the QD gate. The dc gate voltage is adjusted such that the device operates at the point of maximum slope in the phase response. This operating point is stabilized by a proportional-integral feedback loop with a bandwidth below \SI{1}{\hertz}, which continuously corrects the gate offset voltage. When the probe is positioned far from sample gates with applied voltages, the gate offset required to establish the operating point is typically relatively low at around \SI{-2}{\milli\volt}. Electrostatic potential maps are then obtained by recording this compensating gate voltage throughout the scan area and converting it using the lever arm of the QD gate to the QD, $\alpha = C_{\mathrm{G}}/C_{\mathrm{\Sigma}}$.

To remove the background arising from work function differences between the sample and the QD, we first record a reference scan with no voltages applied to the sample gates. In a second measurement scan, we apply a voltage of \SI{-0.5}{\volt} (\SI{0.5}{\volt}) to sample gates G1 and G3 (G2 and G4). The final image is obtained by subtracting the reference scan from the corresponding measurement scan, while additionally a constant offset is subtracted from both datasets in order to set the potential scale to zero. We repeat this procedure at three different scan heights separated by \SI{100}{\nano\meter}. The resulting potential map for the lowest scan height, denoted by $\Delta Z = 0$, is shown in \cref{fig:figure_2_gate_scans}\textbf{c}.

We compare these measurements to COMSOL simulations of the electrostatic potential of the given sample gate geometry. \cref{fig:figure_2_gate_scans}\textbf{d} shows a plane of the simulated data, taken at a height of \SI{100}{\nano\meter} above the top surface of the gates. In \cref{fig:figure_2_gate_scans}\textbf{e}, we compare linecuts of measured data to corresponding linecuts from the simulation, extracted for three different scan heights. Comparison with the simulations yields an estimated scan height of \SI{170}{\nano\meter} between the QD and the sample surface for the lowest scan plane. The relative height differences of the other two scans are given with respect to this position and are determined using laser interferometry. Accounting for the lithographic height of the gates (\SI{75}{\nano\meter}), the QD operates \SI{95}{\nano\meter} above the lithographically defined structures.
\begin{figure}[t]
	\centering
	\includegraphics[width=\linewidth]{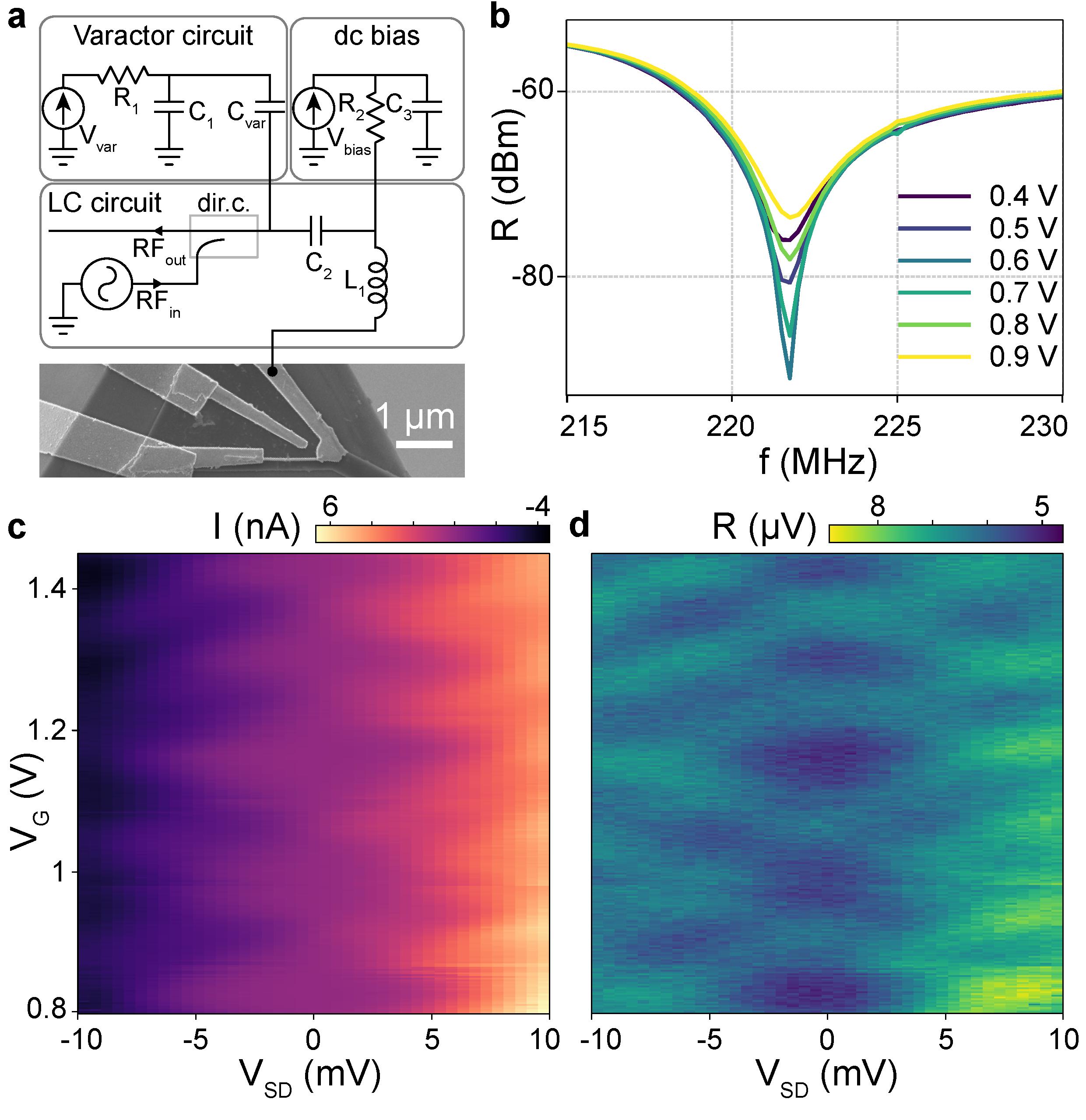}
	\caption{\textbf{Radio-frequency reflectometry read-out of the scanning NW QD.} \textbf{a} Circuit diagram of the rf reflectometry setup. The source contact of the QD is connected to an LC resonator formed by the inductor $L_1$ and the parasitic capacitance to ground. A signal generator provides an oscillating voltage that passes through a directional coupler, indicated by the light gray box, and a coupling capacitor $C_2$. A voltage tunable capacitance $C_{\mathrm{var}}$ is used to match the impedance of the resonant circuit to the transmission line. A bias-tee allows application of a dc voltage to the source electrode.  \textbf{b} Magnitude $R$ of the demodulated reflected signal as a function of frequency for different varactor voltages. The best matching condition is obtained for a varactor voltage of approximately \SI{0.6}{\volt}. \textbf{c} dc current and \textbf{d} magnitude $R$ of the demodulated reflected signal, as a function of source-drain voltage $V_{\mathrm{SD}}$ and QD gate voltage $V_{\mathrm{G}}$. Both measurements exhibit the characteristic Coulomb diamond pattern.}
	\label{fig:rfr}
\end{figure}
The linecuts in \cref{fig:figure_2_gate_scans}\textbf{e} exhibit an overall quantitative agreement, although clear deviations are observable. In the central region, particularly for $\Delta Z = 0$, the experimental potential is lower than the simulated version. Additionally, the experimental traces diverge from the simulated profiles at the edges of the linecuts, an effect most pronounced in the uppermost linecut (blue). These discrepancies likely stem from three primary factors. First, the COMSOL simulations do not account for the physical presence of the probe. The model omits the voltages applied to the QD  gate and source electrodes, as well as the electric screening of the tip volume, which locally distorts the electric field. Because the cantilever continuously changes position during spatial mapping, capturing these tip-induced effects would require resimulation for each pixel. Second, the simple COMSOL model does not take into account the finite size of the sensor QD and consequentially convolution and averaging of the local potential over the sensing volume is not taken into account. Third, low-frequency drifts in the QD operating point, occurring during individual scans, are imprinted on the final image. This dc methodology thus lacks a mechanism for continuous, reliable background subtraction. 
To mitigate such low-frequency drift and improve the signal-to-noise ratio, we tested an ac lock-in detection scheme. By biasing neighboring gates with an out-of-phase ac voltage, this method is insensitive to static dc offsets, thereby eliminating the need for background subtraction. A spatial map of the demodulated phase for this geometry is provided in Appendix V. Although this ac approach yields stable, high-contrast imaging, the extracted data presently remain qualitative due to residual crosstalk and nonlinearities in the measurement chain.

\section*{Radiofrequency reflectometry}
We next demonstrate the integration of rf reflectometry read-out of the scanning SET. Conventional dc current readout of scanning SETs is typically limited to bandwidths of a few tens of kHz, set by the parasitic capacitance of the wiring required to connect the sensor to room-temperature electronics\cite{vigneauProbingQuantumDevices2023}. Read-out via rf reflectometry circumvents this limitation by embedding the sensor in a resonant impedance matching circuit, routinely elevating read-out bandwidths to the MHz regime for fixed, non-scanning, charge sensors\cite{burkardSemiconductorSpinQubits2023}. Bringing this technique to a scanning platform could in principle allow the spatially-resolved detection of fast, local charge dynamics. This could enable, for instance, the measurement of two-level-system defects or other sources of charge noise on timescales within the coherence times of typical solid-state qubits; a capability inaccessible to dc-based scanning charge sensing.

We implement rf reflectometry read-out of our scanning NW QD using an impedance matching circuit mounted on a PCB placed in close vicinity to the cantilever, as shown in \cref{fig:rfr}\textbf{a}. The circuit comprises an LC-resonator connected to the QD source contact via wire bonds from the PCB to the corresponding bondpad on the cantilever chip, a varactor-based circuit to tune the matching\cite{aresSensitiveRadioFrequencyMeasurements2016a}, and a bias-tee. The radio-frequency signal $\mathrm{RF_{in}}$, with a frequency of \SI{221.75}{\mega\hertz}, is passed from the room-temperature microwave source to the impedance matching circuit via a directional coupler, mounted inside the cryostat. The signal is then reflected from the resonant circuit, with a magnitude determined by the QD impedance and the capacitance $C_{\mathrm{var}}$, amplified at \SI{4.2}{\kelvin} using a high-electron-mobility transistor amplifier, and demodulated at room temperature using a lock-in amplifier. The bias-tee provides a quasi-static voltage offset to the source contact of the QD, enabling simultaneous rf and dc read-out. For more information on the reflectometry setup and PCB, see Appendix II.

In \cref{fig:rfr}\textbf{b}, the magnitude of the demodulated signal is shown as a function of frequency for different varactor voltages $V_{\mathrm{var}}$. By tuning the capacitance $C_{\mathrm{var}}$, the varactor enables in situ adjustment of the impedance matching. \Cref{fig:rfr}\textbf{b} indicates that near optimal matching is achieved for $V_{\mathrm{var}} \approx \SI{0.6}{\volt}$, for which case we find a bandwidth of \SI{290}{\kilo\hertz} corresponding to the full-width at half-maximum of the resonance at \SI{221.75}{\mega\hertz}. Using this configuration, we now record Coulomb diamond measurements with rf-reflectometry. The corresponding dc current map and the magnitude $R$ of the demodulated reflected signal are shown in Figs.\ref{fig:rfr}\textbf{c} and \textbf{d}, respectively.

We perform sideband modulation measurements to quantify the performance of this initial implementation of rf readout, yielding an experimental voltage sensitivity of \SI{3.9e-3}{\volt\per\sqrthz} and a corresponding charge sensitivity of \SI{55e-3}{\elementarycharge\per\sqrthz} at \SI{221.75}{\mega\hertz}. Detailed sensitivity calculations are provided in Appendix III. While these values are modest compared to state-of-the-art rf-SETs/QDs\cite{vigneauProbingQuantumDevices2023}, the reduced sensitivity and bandwidth are most likely a consequence of the unoptimized high-frequency environment of the cantilever. Specifically, it is probable that a large fraction of the incident rf power is dissipated by the finite conductivity of the highly doped Si cantilever substrate\cite{schollhornAttenuationMechanismsAluminum2003} and the large loss tangent of the AlO$_x$ dielectric layer\cite{pappasTwoLevelSystem2011}. Furthermore, the leads connecting the QD to the LC matching circuit on the PCB are simple single-line traces and wire bonds rather than impedance-matched transmission lines. These factors reduce the power transfer to the QD, suppressing its participation in the circuit. We expect that transitioning to cantilever substrates with lower doping and implementing impedance-matched signal routing will substantially boost the sensitivity in future iterations, enabling rf read-out during scanning probe measurements.
\section*{Conclusions}
In conclusion, we have established a fabrication protocol for the reliable placement and contacting of a nanowire quantum dot at the apex of a NC-AFM cantilever. The resulting probe maintains NC-AFM functionality, providing the robust height control required to reliably navigate complex surface topographies. We demonstrate the probe's functionality as a scanning electrometer, by mapping the electrostatic potential generated by four nanoscale gate electrodes. These measurements show quantitative agreement with electrostatic simulations, which further reveal a low operating height of approximately \SI{100}{\nano\meter} for the active sensor element. Finally, we demonstrate that this platform is compatible with rf reflectometry read-out by acquiring Coulomb diamond measurements via the reflected signal. Together, these results establish a pathway toward robust, high-bandwidth scanning electrometry of complex nanostructures.
\section*{Acknowledgements}
We thank C. Sch\"onenberger for fruitful discussions, as well as S. Martin and M. Steinacher for technical support. We acknowledge the support of the Swiss National Science Foundation via the NCCR SPIN, as well as the Canton Aargau.

\section*{Appendix I: Fabrication procedure}
\begin{figure} [b]
	\centering
	\includegraphics[width=\linewidth]{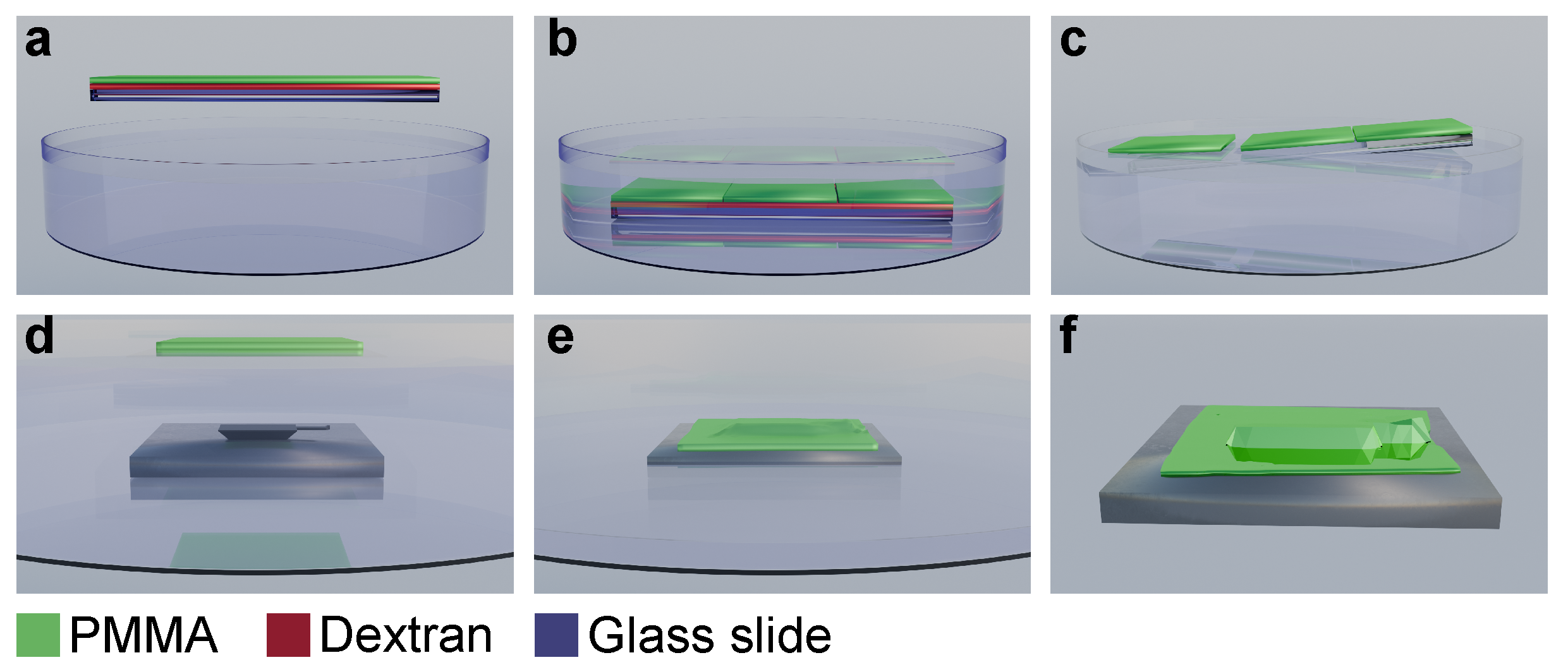}
	\caption{Schematic of the refined floating PMMA coating procedure. \textbf{a} A glass slide is spin-coated with a sacrificial dextran layer and PMMA. \textbf{b} The film is cut into rectangular pieces ($\sim\SI{1}{\centi\meter}\times\SI{0.5}{\centi\meter}$) and submerged in DI H$_2$O where the dextran layer dissolves. \textbf{c} The PMMA film delaminates and floats on the water surface upon reinsertion of the slide at a shallow angle. \textbf{d} A single floating PMMA flake is transferred to a DI H$_2$O-filled Petri dish, immobilized with a glass needle, and positioned above a submerged cantilever fixed to an Al plate. \textbf{e, f} The water is slowly drained, allowing the PMMA film to gently settle onto the cantilever. Excess water is removed under gentle vacuum and a final bake evaporates residual moisture, yielding a uniformly coated lever.}
	\label{figure_3_floating_pmma}
\end{figure}
\subsection{Cantilever preparation}
The probe is based on a commercially available NC-AFM cantilever (Nanosensors
ATEC-NC). A flat mesa, which later hosts the nanowire device, is milled at the
cantilever apex by focused ion beam. Since the cantilever substrate consists
of highly doped silicon, an insulating layer is required to electrically
isolate the device from the cantilever chip; to this end, $\sim$\SI{60}{\nano\meter}
of AlO$_x$ is deposited by atomic layer deposition. Subsequently, five Ti/Au
(\SI{5}{\nano\meter}/\SI{120}{\nano\meter}) bond pads are evaporated onto the
cantilever chip through a silicon shadow mask. A scanning electron micrograph
of a cantilever prepared in this way is shown in Fig.~\ref{fig:figure_1}\textbf{a}.

\subsection{Device fabrication on silicon nitride membranes}
NW QD device fabrication is done in batches, where several devices are fabricated in parallel. The nanowires have a total length of \SI{1.9}{\micro\meter}, a diameter of \SI{50}{\nano\meter}, and have two InP axial tunnel barriers (width $\sim$\SI{5.5}{\nano\meter}) spaced by \SI{19}{\nano\meter} and placed approximately \SI{440}{\nano\meter} from one end of the nanowires. The nanowire devices are prepared on suspended silicon nitride membranes. The membranes are fabricated from a double-side-polished undoped $\langle100\rangle$ silicon wafer, onto which a \SI{50}{\nano\meter} thick silicon
nitride layer is deposited on the front side and a \SI{300}{\nano\meter} thick
layer on the back side by plasma-enhanced chemical vapor deposition. Square
openings are defined on the back side by photolithography and transferred
through the silicon nitride layer by reactive ion etching. The exposed silicon
is removed by anisotropic wet etching in KOH, yielding suspended
\SI{50}{\nano\meter} thick silicon nitride membranes.

After definition and deposition of alignment markers, nanowires are dispersed
onto the membranes by micromanipulation. Source and drain contacts as well as
a plunger gate adjacent to the QD are patterned by electron beam lithography.
The native oxide on the nanowires is removed by in situ argon ion beam etching
prior to metallization (\SI{5}{\nano\meter} Ti, \SI{20}{\nano\meter} Pd,
\SI{50}{\nano\meter} Au). An additional electron beam lithography step
followed by reactive ion etching yields prefabricated devices attached to the
surrounding membrane only by narrow silicon nitride scaffolding beams, as
depicted in Fig.~\ref{fig:figure_1}\textbf{b}. To protect the device during the
subsequent transfer, the residual resist is intentionally left on the
structure after etching.

\subsection{Device transfer and contacting}
The prefabricated device is transferred to the mesa of the prepared cantilever
using a micromanipulator. During this step, the sacrificial scaffolding beams
are broken with a glass needle, yielding the assembly shown in
Fig.~\ref{fig:figure_1}\textbf{c}. Residual resist on the SiN support is removed in an
oxygen plasma.

Since standard spin coating is not applicable to AFM cantilevers, an
alternative resist coating strategy is required for electron beam lithography
on the assembled probe. The levers are therefore coated using a refined
version of the floating PMMA technique described in our previous work
\cite{forrerElectronbeamLithographyNanostructures2023}, illustrated in Fig.\ref{figure_3_floating_pmma}.
Electron beam lithography followed by evaporation of \SI{5}{\nano\meter} Ti
and \SI{70}{\nano\meter} Au establishes electrical contact between the device
and the bond pads on the cantilever chip. Finally, \SI{20}{\nano\meter} of Au
is evaporated onto the backside of the cantilever to enhance the optical
reflectance for interferometric readout, and an additional
\SI{30}{\nano\meter} AlO$_x$ layer is deposited by atomic layer deposition.
This yields the contacted nanowire QD device integrated at the cantilever
apex shown in Fig.~\ref{fig:figure_1}\textbf{d}.

\section*{Appendix II: Cantilever printed circuit board}
\begin{figure}[h]
	\centering
	\includegraphics[width=\linewidth]{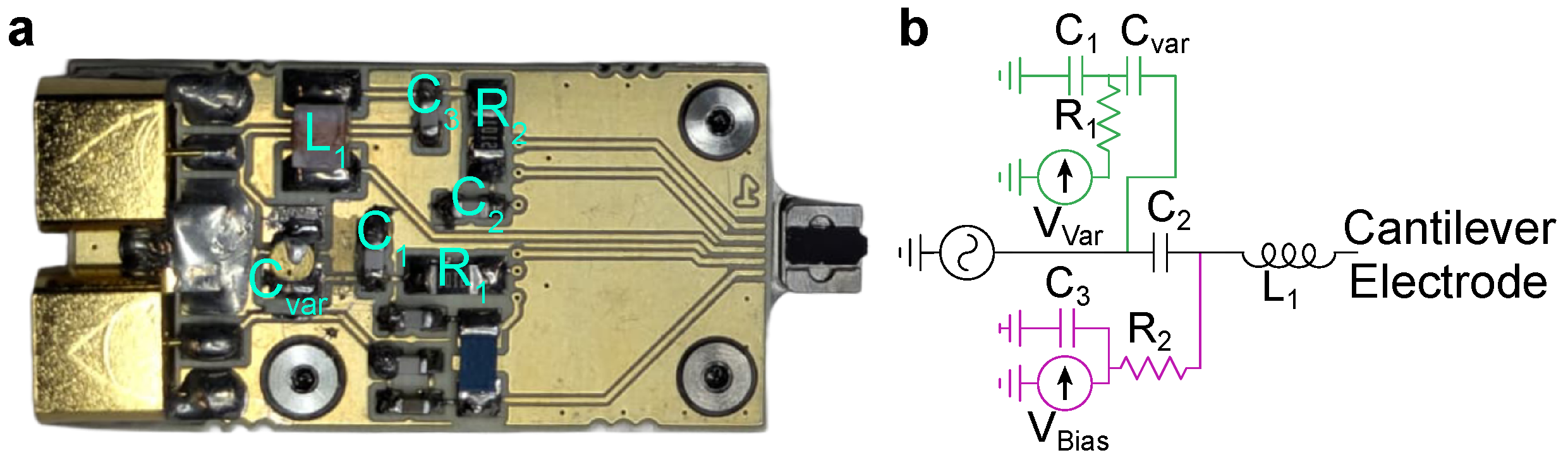}
	\caption{Cantilever printed circuit board. \textbf{a} Optical image of cantilever side of the printed circuit board, with components of LC-circuit and bias-tee indicated. On the right side of the printed circuit board, the cantilever is glued on a pedestal and its electrodes are connected via wire bonds. The components at the bottom of the printed circuit board are not used in this work. \textbf{b} Diagram of the circuit mounted on the printed circuit board.}
	\label{figure_S_PCB}
\end{figure}
\begin{table}[h]
	\centering
	\sisetup{separate-uncertainty=true, multi-part-units=single} 
	\begin{tabular}{ll}
		\toprule
		\textbf{Component} & \textbf{Value} \\
		\midrule
		R$_1$              & \SI{5}{\kilo\ohm} \\
		R$_2$              & \SI{5}{\kilo\ohm} \\
		C$_1$              & \SI{330}{\pico\farad} \\
		C$_2$              & \SI{100}{\pico\farad} \\
		C$_3$ 				& \SI{87}{\pico\farad} \\
		L$_1$              & \SI{470}{\nano\henry} \\
		C$_{var}$          & Macom MA46H204-1056 \\
		\bottomrule
	\end{tabular}
	\caption{Component values of the lever pcb used for the measurements presented in this work.}
	
	\label{tab:rf-pcb}
\end{table}

\section*{Appendix III: QD parameter extraction}
\begin{figure}[h]
	\centering
	\includegraphics[width=\linewidth]{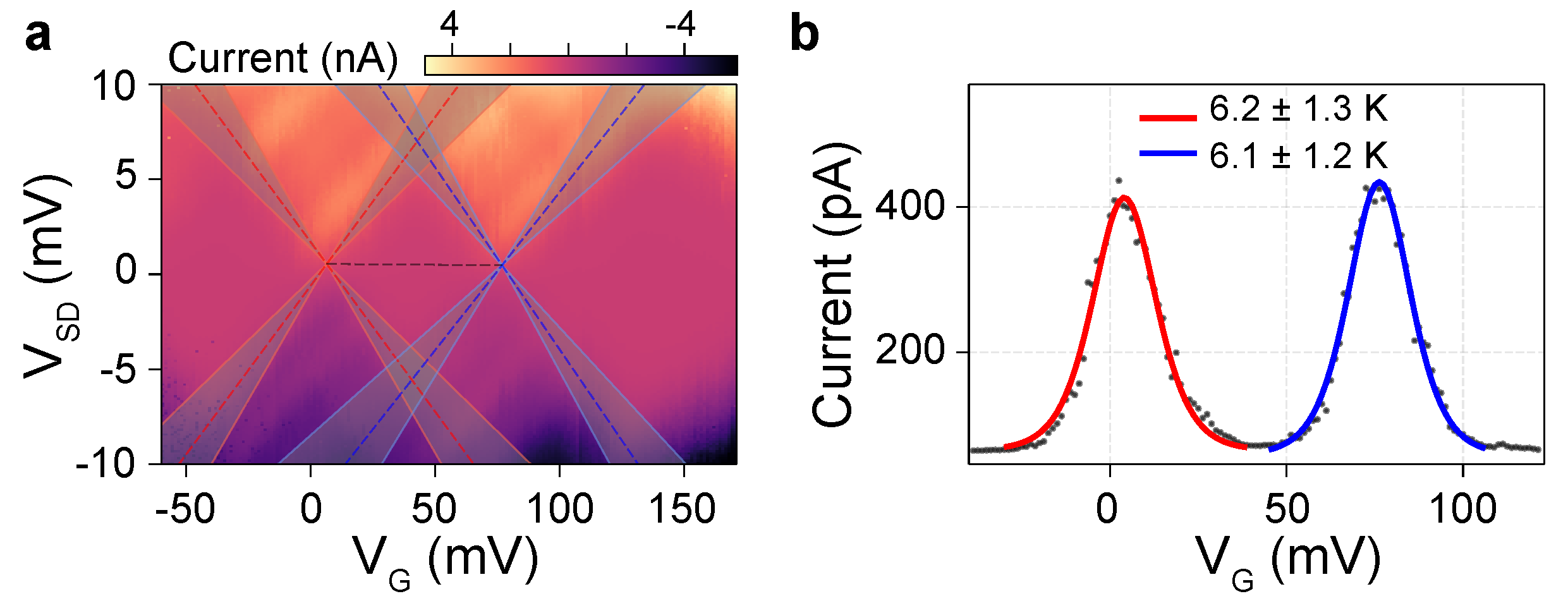}
	\caption{Coulomb diamond analysis and electron temperature extraction. \textbf{a} DC current map measured as a function of gate voltage ($V_\mathrm{G}$) and source-drain bias ($V_\mathrm{SD}$). Dashed lines highlight the diamond edges used to extract the gate lever arm $\alpha$ for two distinct charge states (red and blue). The corresponding solid lines and shaded regions represent the error margins of the linear estimate. A black dashed line connects the diamond vertices, yielding the gate voltage spacing $\Delta V_\mathrm{G}$. \textbf{b} Current trace of Coulomb oscillations (black dots) measured as a function of $V_\mathrm{G}$ at a fixed source-drain bias of \SI{0.5}{\milli\volt}. Both peaks are fitted with equation for temperature-broadened Coulomb oscillations \cite{beenakkerTheoryCoulombblockadeOscillations1991} (solid lines). Using the lever arm $\alpha$ extracted in \textbf{a}, these fits determine the electron temperature $T_e$ of the system.}
	\label{fig:LeverArm_E-Temperature}
\end{figure}

\begin{table}[htbp]
	\centering
	\sisetup{separate-uncertainty=true, multi-part-units=single} 
	\begin{tabular}{llc}
		\toprule
		\textbf{Parameter} & \textbf{Symbol} & \textbf{Extracted Value} \\
		\midrule
		Gate lever arm     & \(\alpha\)       & \SI{8.9 +- 1.8}{\percent} \\
		Gate capacitance   & \(C_\mathrm{G}\) & \SI{2.3 +- 0.2}{\atto\farad} \\
		Source capacitance & \(C_\mathrm{S}\) & \SI{11.9 +- 3.2}{\atto\farad} \\
		Drain capacitance  & \(C_\mathrm{D}\) & \SI{11.5 +- 6.4}{\atto\farad} \\
		Total capacitance  & \(C_{\Sigma}\)      & \SI{25.7 +- 5.5}{\atto\farad} \\
		Charging energy    & \(E_\mathrm{C}\) & \SI{6.2 +- 1.3}{\milli\electronvolt} \\
		\bottomrule
	\end{tabular}
	\caption{Summary of extracted QD parameters in the constant interaction model \cite{hansonSpinsFewelectronQuantum2007, kouwenhovenFewelectronQuantumDots2001}. The reported errors are propagated from the estimation of the Coulomb diamond slopes and the peak-to-peak spacing $\Delta V_\mathrm{G}$.}
	\label{tab:ci-model-parameters}
\end{table}

\subsection{Sensitivity Calculation}
\subsubsection{Sensitivity dc read-out}
The charge sensitivity is calculated via \cite{vigneauProbingQuantumDevices2023, zimmerliNoiseCoulombBlockade1992}
\begin{equation}
	\delta Q(f) = \frac{i_n(f)}{\left|\frac{\partial I}{\partial Q}\right|} = C_G\frac{i_n(f)}{\left|\frac{\partial I}{\partial V_G}\right|}
\end{equation}
where $i_n(f)$ is the current noise density, $\partial I / \partial Q$ is the response of the QD current to a change in the induced charge $Q$, and $C_{\mathrm{G}}$ is the QD gate capacitance. In our setup, the current is converted to a voltage via an I to V converter with a gain of \SI[retain-unity-mantissa=false]{1e8}{\volt\per\ampere} and a \SI{1}{\mega\hertz} bandwidth. The charge sensitivity can thus be expressed in terms of the measured voltage noise density $e_n(f)$ as    
\begin{equation}
	\delta Q(f) = C_{\mathrm{G}} \frac{e_n(f)}{\left|\frac{\partial V_{\mathrm{out}}}{\partial V_{\mathrm{G}}}\right|}
\end{equation}

The voltage noise density $e_n(f)$ is extracted from a power spectral density (PSD) measurement recorded with the cantilever retracted from the sample surface, the piezoelectric actuator disabled, the readout laser turned off, and an external magnetic field of \SI{1.5}{\tesla} applied. The magnetic field reduces mechanical vibrations of the scanning probe via eddy current damping.

The system gain $\left|\partial V_{\mathrm{out}} / \partial V_{\mathrm{G}}\right|$ is determined from the lock-in measurement at \SI{1.25}{\kilo\hertz}. Dividing the maximum lock-in amplitude (after subtracting a constant background offset) by the applied \SI{2.2}{\milli\volt} rms gate excitation yields a maximum experimental system gain of $\approx \SI{0.33}{\volt\per\volt}$. Inserting this gain into the equation above, alongside the measured voltage noise at \SI{1.25}{\kilo\hertz} and the gate capacitance extracted from Fig.~\ref{fig:LeverArm_E-Temperature} (see \cref{tab:ci-model-parameters}), yields a charge sensitivity of \SI{425e-6}{\elementarycharge\per\sqrthz}. The corresponding voltage sensitivity, given by $\delta V = \delta Q / C_{\mathrm{G}}$, is \SI{30e-6}{\volt\per\sqrthz}.

Compared to the system gain extracted from the dc slope of the Coulomb oscillations, this ac response is considerably reduced. We attribute this signal attenuation to the low-pass RC filter formed by the relatively high QD resistance at the typical operating point ($\sim$\SI{1.5}{\mega\ohm}) and the parasitic capacitance of the readout wiring.

\subsubsection{Sensitivity rf-reflectometry read-out}
To determine the rf sensitivity, the QD is biased to the steepest flank of a Coulomb oscillation, and this operating point is stabilized by a sub-Hertz PI loop. Subsequently, the varactor voltage is tuned to minimize the reflected signal in the resonance dip during a frequency sweep. These optimal matching conditions are maintained for the remainder of the measurement. Following Vigneau \textit{et al.} \cite{vigneauProbingQuantumDevices2023}, the sensitivity can be extracted from the signal-to-noise ratio (SNR) of a gate-voltage-induced modulation sideband:
\begin{equation}
	\delta Q = \frac{\delta q_{\mathrm{rms}}}{\sqrt{2 \Delta_f} \times 10^{\mathrm{SNR_{\mathrm{dB}}}/20}}
\end{equation}
where $\delta q_{\mathrm{rms}}$ is the calibrated rms charge modulation applied to the QD, and $\Delta_f$ is the resolution bandwidth. 
In our experiment, we apply a modulation of $\delta q_{\mathrm{rms}} \approx \SI{0.02}{\elementarycharge}$ at $f_{\mathrm{mod}} = \SI{90}{\hertz}$. Because we extract the SNR from the demodulated signal rather than the raw power spectrum, the factor of $\sqrt{2}$ in the denominator is omitted. Furthermore, because the signal is processed using a Hann window, the resolution bandwidth $\Delta_f$ is scaled by a factor of 1.5 to account for the increased equivalent noise bandwidth (ENBW) of the filter \cite{heinzel2002spectrum}. Evaluating this adjusted expression yields a charge sensitivity of \SI{55e-3}{\elementarycharge\per\sqrthz} at a frequency of \SI{221.75}{\mega\hertz}. Converting this value via the gate capacitance yields a corresponding voltage sensitivity of \SI{3.9e-3}{\volt\per\sqrthz}.

\section*{Appendix IV: Scanning probe methods}
We perform all transport characterization, NC-AFM, and scanning SET measurements in a custom-built microscope under high vacuum inside a 4He cryostat with a temperature at the scanning probe of \SI{4.8}{\kelvin}. Piezoelectric walkers and scanners (Attocube ANPx311/LT/HV, ANSxy100lr/LT/HV) are used to move the sample with respect to the fixed cantilever. The cantilever’s flexural motion is monitored by a fiber-optic interferometer, consisting of a \SI{1550}{\nano\meter} diode laser (Toptica), a 99:1 fiber-optic coupler, and a custom-built objective, focusing light onto the cantilever to an \SI{8}{\micro\meter} spot. Interference from the low-finesse Fabry-P\'erot interferometer formed by the cleaved face of the optical fiber and the cantilever shaft is measured by a photoreceiver and used for cantilever displacement measurement. The cavity is stabilized against drift using a PID controller to control the laser temperature. The cantilever and its holder are mounted at a 10 degree angle with respect to the sample surface, such that the end of the cantilever is the first part to come into contact with the sample. For topographic imaging, the cantilever is excited to an amplitude of \SI{5}{\nano\meter} at its fundamental mechanical resonance frequency of \SI{331}{\kilo\hertz} by a piezoelectric actuator driven by a phase-locked loop (Nanonis BP5e).

\section*{Appendix V: AC Potential Map}
\begin{figure}[h]
	\centering
	\includegraphics[width=\linewidth]{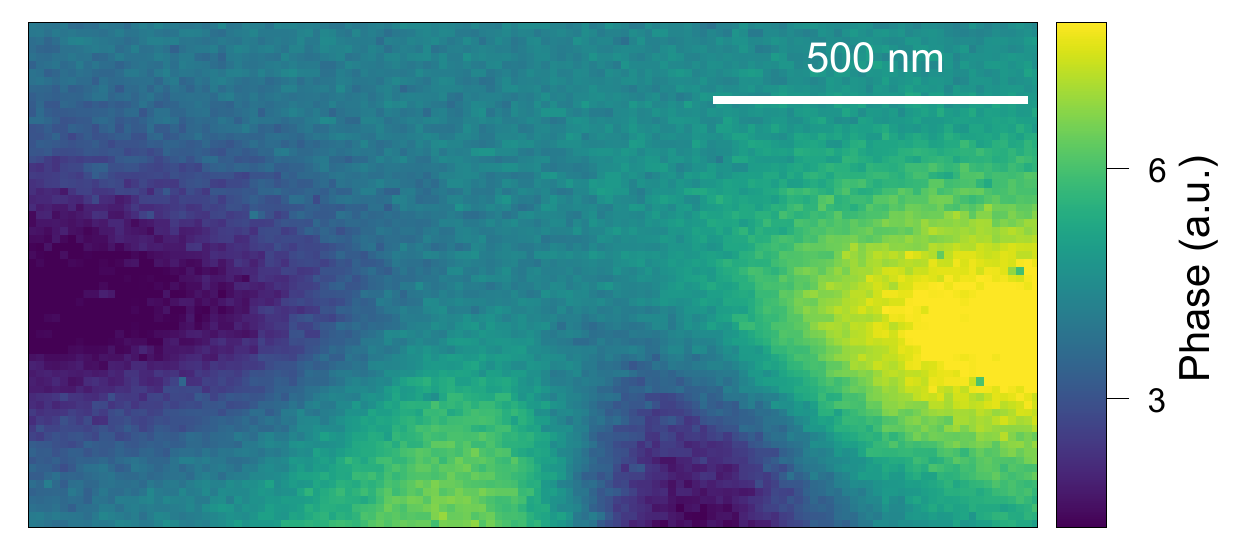}
	\caption{ac lock-in imaging. Qualitative visualization of the electrostatic potential modulation generated by the sample electrodes, acquired using an ac excitation of \SI{50}{\milli\volt} at \SI{227}{\hertz} with a \SI{180}{\degree} phase shift between neighboring gates. The map shows the demodulated phase signal detected by the probe QD.}
	\label{fig:AC-GateScan}
\end{figure}
To partially suppress $1/f$ noise and narrow the measurement bandwidth for an increased signal-to-noise ratio, we implemented an ac lock-in mapping scheme. Furthermore, because this technique is insensitive to static dc offsets, it eliminates the need for background reference scans. Analogous to the dc methodology, a sub-Hertz PI loop stabilizes the QD operating point on the flank of a Coulomb oscillation. Neighboring sample gates are driven out-of-phase using a \SI{50}{\milli\volt} excitation at \SI{227}{\hertz}. The AC potential at any given probe position is a linear superposition of the fields generated by the individual sample electrodes. Combined with a constant background phase offset caused by capacitive line crosstalk, this out-of-phase excitation causes the demodulated lock-in phase to transition between opposite polarities above neighboring gates. As shown in \cref{fig:AC-GateScan}, this dynamic measurement yields a spatial map that resolves the same underlying gate geometry as the DC scans. However, due to the aforementioned residual crosstalk and nonlinearities in the readout chain, these AC maps presently remain qualitative.
\bibliographystyle{apsrev4-2}
\bibliography{2026SNWQDPaper}
\end{document}